\documentclass{PoS}

\usepackage{graphicx}

\newcommand{\note}[1]{}                    

\newcommand{\mnote}[1]{}                   

\newcommand{\sixth}{\mbox{\small $\frac{1}{6}$}}         
\newcommand{\half}{\mbox{\small $\frac{1}{2}$}}          

\def\lsim{\mathrel{\rlap{\lower4pt\hbox{\hskip1pt$\sim$}}
    \raise1pt\hbox{$<$}}}                
\def\gsim{\mathrel{\rlap{\lower4pt\hbox{\hskip1pt$\sim$}}
    \raise1pt\hbox{$>$}}}                

\title{
\vspace*{-1.25cm}
\begin{minipage}{\textwidth}
\begin{flushright}
\texttt{\footnotesize
PoS(LATTICE 2013)249 \\%
ADP-13-23/T843       \\%
DESY 13-220          \\%
Edinburgh 2013/30    \\%
Liverpool LTH 993    \\%
}
\end{flushright}
\end{minipage}\\[15pt]
\vspace*{+1.25cm}
       SU(3) flavour symmetry breaking and charmed states}

\ShortTitle{SU(3) flavour symmetry breaking and $\ldots$}

\author{\speaker{R. Horsley}$^{\,a}$,
        J. Najjar$^b$,
        Y. Nakamura$^c$,
        H. Perlt$^d$,
        D. Pleiter$^{eb}$, 
        P.~E.~L. Rakow$^f$,
        G. Schierholz$^g$,
        A. Schiller$^d$,
        H. St\"uben$^h$
        and J.~M. Zanotti$^i$ \\
        \llap{$^a$} School of Physics and Astronomy,
                    University of Edinburgh,
                    Edinburgh EH9 3JZ, UK \\
        \llap{$^b$} Institut f\"ur Theoretische Physik,
                    Universit\"at Regensburg, 93040 Regensburg, Germany \\
        \llap{$^c$} RIKEN Advanced Institute for Computational Science,
                    Kobe, Hyogo 650-0047, Japan \\
        \llap{$^d$} Institut f\"ur Theoretische Physik,
                    Universit\"at Leipzig, 04109 Leipzig, Germany \\
        \llap{$^e$} JSC, Forschungszentrum J\"ulich,
                    52425 J\"ulich, Germany \\
        \llap{$^f$} Theoretical Physics Division,
                    Department of Mathematical Sciences,
                    University of Liverpool,
                    Liverpool L69 3BX, UK \\
        \llap{$^g$} Deutsches Elektronen-Synchrotron DESY,
                    22603 Hamburg, Germany \\
        \llap{$^h$} Regionales Rechenzentrum, Universit\"at Hamburg,
                    20146 Hamburg, Germany \\
        \llap{$^i$} CSSM, School of Chemistry and Physics,
                    University of Adelaide, Adelaide SA 5005, Australia \\
        E-mail: \email{rhorsley@ph.ed.ac.uk} }

\author{QCDSF-UKQCD Collaborations}

\abstract{
   By extending the SU(3) flavour symmetry breaking expansion
   from up, down and strange sea quark masses to partially quenched
   valence quark masses we propose a method to determine charmed
   quark hadron masses including possible QCD isospin breaking
   effects. Initial results for some open charmed pseudoscalar
   meson states and singly and doubly charmed baryon
   states are encouraging and demonstrate the potential of the procedure.
   Essential for the method is the determination of the scale 
   using singlet quantities, and to this end we also give here
   a preliminary estimation of the recently introduced
   Wilson flow scales.}

\FullConference{31st International Symposium on Lattice Field Theory 
                - LATTICE 2013\\
		July 29 - August 3, 2013\\
		Mainz, Germany}

\begin{document}


\section{Motivation and Strategy}


At present there is considerable interest in open charm masses.
While much is known about $C=1$ charmed meson masses and singly
charmed baryon masses, the situation is far less clear for
doubly charmed quark baryons (i.e.\ $C=2$ or $ccq$ with $q =u$,
$d$, $s$). Here many masses are unknown, but as stable states
in QCD they must exist (presently only some candidate states are
seen by the SELEX Collaboration \cite{selex02a}, but not by BaBar
\cite{aubert06a}, or BELLE \cite{chistov06a}).
Also there are possible relations to tetraquark states.
For example in the $n_c \to \infty$, \cite{weinberg13a},
$m_c \to \infty$ limit we expect the relation
$M(cc\overline{u}\overline{d}) \approx M_{\Xi_{cc}^{++}} + M_{\Lambda_c^+}
 -M_{D^0} - (M_{D^{*+}}+M_{D^0})/4$, \cite{karliner13a}.

The charm quark, $c$ is considerably heavier than the up $u$,
down $d$ and strange $s$ quarks, which has hampered its
direct simulation using lattice QCD. However as the available
lattice spacings become finer, this is becoming less of an obstacle.
The sea quarks in present day $n_f = 2+1$ flavour dynamical
lattice simulations consist of two mass degenerate (i.e.\ $m_u = m_d$)
light flavours $u$, $d$ and a heavier flavour $s$.
Their masses are typically larger than the `physical' masses
necessary to reproduce the experimental spectrum. How can we
usefully approach the `physical' $u$, $d$, $s$ quark masses?
One possibility suggested in \cite{bietenholz11a} is to
consider an $SU(3)$ flavour breaking expansion from a point
$m_0$ on the flavour symmetric line keeping the average quark mass
$\overline{m} = (m_u+m_d+m_s)/3$ constant ($= m_0$).
This not only significantly reduces the number of expansion
coefficients allowed, but the expansion coefficients remain
the same whether we consider $m_u \not= m_d$ or $m_u = m_d$.
Thus we can also find the pure QCD contribution to isospin
breaking effects with just one $n_f = 2+1$ numerical simulation.

The $SU(3)$ flavour breaking expansion can also be extended
to valence quark masses, i.e.\ the quarks making up the meson
or baryon have not necessarily the same mass as the sea quarks.
The valence quarks are called `Partially Quenched' or PQ quarks
in distinction to the sea or dynamical sea quarks.
We call the `Unitary Limit' when the masses of the valence
quarks coincide with the sea quarks. PQ determinations
have the advantage of being cheap compared to
dynamical simulations and including them allows a better
determination of the expansion coefficients over a wider range
of quark masses. (This was the strategy pursued in
\cite{horsley12a}.)
In addition because the charm quark, $c$ is much heavier than the $u$,
$d$ and $s$ quarks, it contributes little to the dynamics
of the sea and so we can regard the charm quark as a PQ quark.


\section{Method}
\label{method}


We presently only consider hadrons which lie on the outer ring
of their associated multiplet and not the central hadrons.
(So we need not consider either any mixing or numerical
evaluation of quark--line disconnected correlation functions.)
The $SU(3)$ flavour symmetry breaking expansion for the pseudoscalar
mesons with valence quarks $a$ and $b$ up to cubic or NNLO terms in the
quarks masses is given by
 \begin{eqnarray}
         M^2(a\overline{b})
            &=& M^2_{0\pi} + \alpha(\delta\mu_a + \delta\mu_b)
                                                         \nonumber   \\
            & &
            + \beta_0\sixth(\delta m_u^2 + \delta m_d^2 + \delta m_s^2)
            + \beta_1(\delta\mu_a^2 + \delta\mu_b^2)
            + \beta_2(\delta\mu_a - \delta\mu_b)^2
                                                         \nonumber   \\
            & &
            + \gamma_0\delta m_u\delta m_d\delta m_s
            + \gamma_1(\delta\mu_a + \delta\mu_b)
                       (\delta m_u^2 + \delta m_d^2 + \delta m_s^2)
                                                         \nonumber   \\
            & &
            + \gamma_2(\delta\mu_a + \delta\mu_b)^3
            + \gamma_3(\delta\mu_a + \delta\mu_b)
                              (\delta\mu_a - \delta\mu_b)^2 \,,
\label{M2ps_expan}
\end{eqnarray}
with
$\delta\mu_q = \mu_q - \overline{m}$, $q = a, b, \ldots \in \{u, d, s, c\}$
being valence quarks of arbitrary mass, $\mu_q$ and
$\delta m_q = m_q - \overline{m}$, $q \in \{u, d, s\}$
being sea quarks. (These have the automatic constraint
$\delta m_u + \delta m_d + \delta m_s = 0$.)
Note that we have some mixed sea/valence mass terms.
The unitary limit occurs when $\delta\mu_q \to \delta m_q$.
The expansion coefficients are $M_{0\pi}^2(\overline{m})$,
$\alpha(\overline{m})$, $\ldots$ so if $\overline{m}$ is held constant
then we have constrained fits to the numerical data.
In particular a $n_f = 2+1$ flavour simulation, when
$\delta m_u = \delta m_d \equiv \delta m_l$
is enough to determine the expansion coefficients.
We now use the PQ (and unitary) data to determine the expansion
coefficients (i.e.\ $\alpha$s, $\beta$s, $\gamma$s).
This in turn leads to a determination of the `physical'
quark masses $\delta m_u^*$, $\delta m_d^*$, $\delta m_s^*$
and $\delta \mu_c^*$ by fitting to e.g.\
$M^{\exp}_{\pi^+}(u\overline{d})$, $M^{\exp}_{K^+}(u\overline{s})$ and
$M^{\exp}_{\eta_c}(c\overline{c})$. We can now describe pseudoscalar
open charm states with the same wavefunction (and hence expansion)
${\cal M} = \overline{q}\gamma_5 c$ ($q = u$, $d$, $s$) i.e.\
$D^0(c\overline{u})$, $D^+(c\overline{d})$ and $D_s^+(c\overline{s})$.
Using $\delta m_u^*$, $\delta m_d^*$, $\delta m_s^*$
and $\delta \mu_c^*$ gives estimates of their physical masses.

Similarly for the baryon octet the same procedure can be applied.
We have the $SU(3)$ flavour symmetry breaking expansion
\begin{eqnarray}
   M^2(aab)
      &=& M_{0N}^2 + A_1(2\delta\mu_a + \delta\mu_b) 
                + A_2(\delta\mu_b - \delta\mu_a)
                                                         \nonumber  \\
      & &       + B_0\sixth(\delta m_u^2 + \delta m_d^2 + \delta m_s^2)
                + B_1(2\delta\mu_a^2+\delta\mu_b^2)
                + B_2(\delta\mu_b^2-\delta\mu_a^2) 
                + B_3(\delta\mu_b-\delta\mu_a)^2
                                                         \nonumber   \\
      & &       + C_0\delta m_u\delta m_d\delta m_s
                + [ C_1(2\delta\mu_a + \delta\mu_b) 
                    + C_2(\delta\mu_b - \delta\mu_a)
                  ](\delta m_u^2 + \delta m_d^2 + \delta m_s^2)
                                                         \nonumber   \\
      & &        + C_3(\delta\mu_a + \delta\mu_b)^3
                 + C_4(\delta\mu_a + \delta\mu_b)^2(\delta\mu_a - \delta\mu_b)
                                                         \nonumber   \\
      & &
                 + C_5(\delta\mu_a + \delta\mu_b)
                                     (\delta\mu_a - \delta\mu_b)^2
                 + C_6(\delta\mu_a - \delta\mu_b)^3 \,,
\label{M2N_expan}
\end{eqnarray}
(so for example $M_p = M(uud)$). Again we use PQ (and unitary)
data to first determine the expansion coefficients (i.e.\ the
$A$s, $B$s, $C$s). We can then describe charm states with the
same nucleon like wavefunction and hence same expansion.
For example for single open charm ($C = 1$) states we have
$\Sigma_c^{++}(uuc)$, $\Sigma_c^0(ddc)$, $\Omega_c^0(ssc)$
which all have the wavefunction ${\cal B} = \epsilon (q^TC\gamma_5c)q$
($q = u$, $d$, $s$) while for double open charm ($C = 2$) states
we have $\Xi_{cc}^{++}(ccu)$, $\Xi_{cc}^+(ccd)$, $\Omega^+_{cc}(ccs)$
which all have the wavefunction ${\cal B} = \epsilon (c^TC\gamma_5q)c$
($q = u$, $d$, $s$). In both cases using $\delta m_u^*$,
$\delta m_d^*$, $\delta m_s^*$ and $\delta \mu_c^*$ gives
estimates of these physical masses.


\section{Lattice details}


We use a tree level Symanzik gluon action and an $O(a)$ improved
clover fermion action, including mild stout smearing \cite{cundy09a}.
Thus the quark mass is given by
\begin{eqnarray}
   m_q = {1 \over 2}
           \left( {1\over \kappa_q} - {1\over \kappa_{0c}} \right) \,,
\end{eqnarray}
where $\kappa_q$ is the hopping parameter, $\kappa_0$ is the
hopping parameter along the symmetric line with $\kappa_{0c}$ being
its chiral limit. We shall consider two lattice spacings
($\beta = 5.50$ on $32^3\times 64$ lattices and preliminary
results for $\beta = 5.80$ on $48^3\times 96$ lattices).

\begin{figure}[h]

\begin{minipage}{0.45\textwidth}

   \begin{center}
      \includegraphics[width=7.00cm]
            {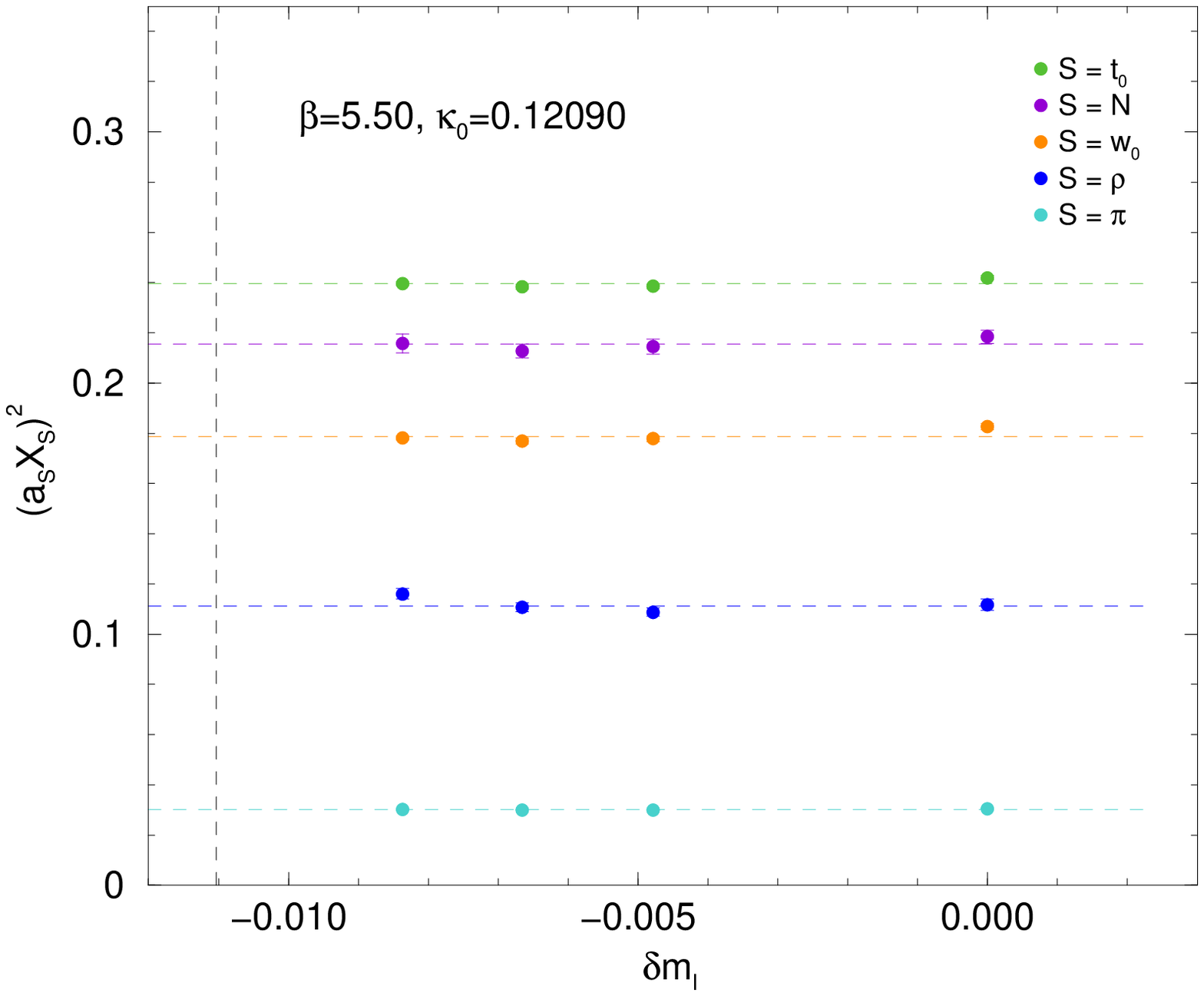}
   \end{center} 

\end{minipage}\hspace*{0.05\textwidth}
\begin{minipage}{0.45\textwidth}

   \begin{center}
      \vspace*{0.75cm}
      \includegraphics[width=7.25cm]
            {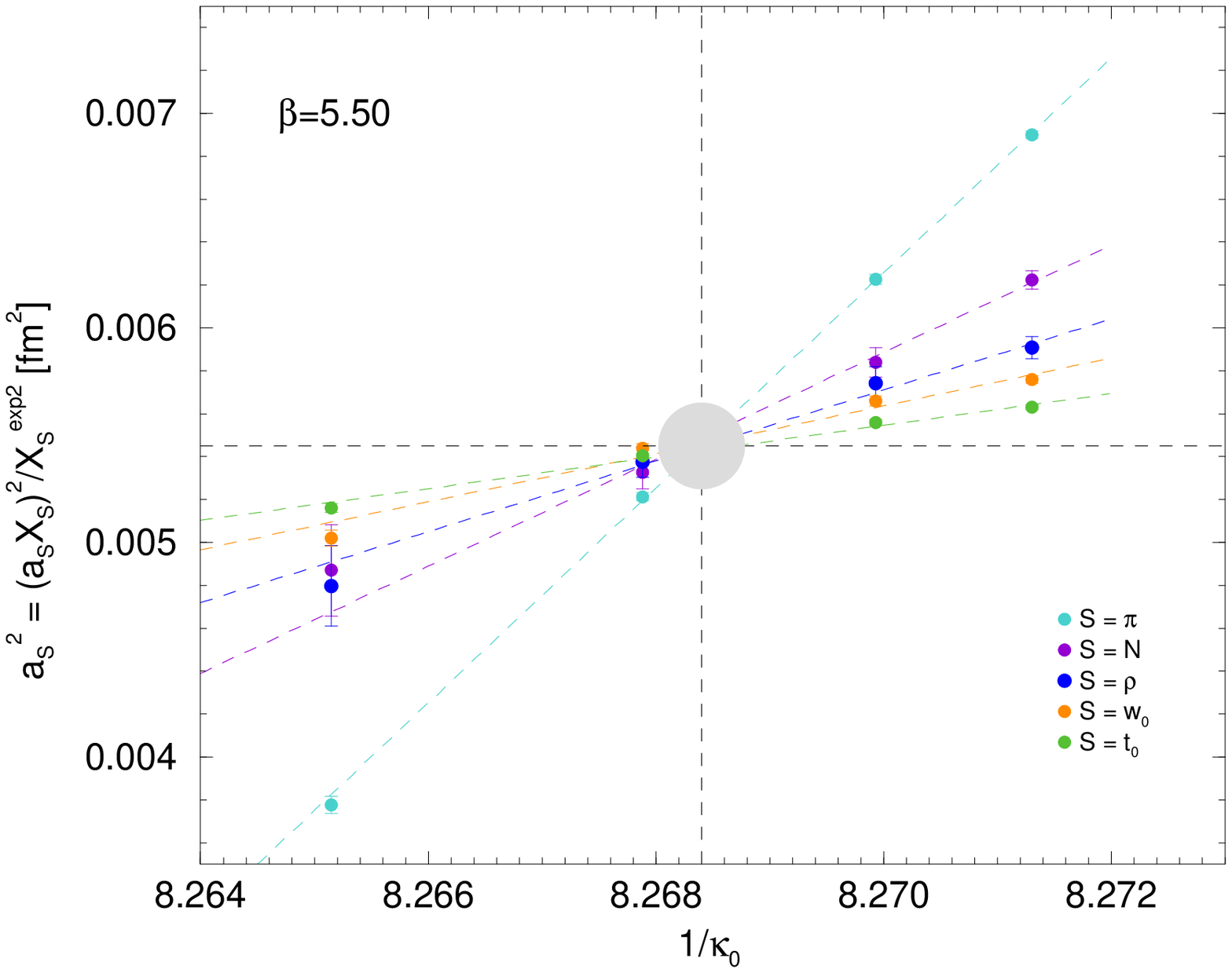}
   \end{center} 

\end{minipage}
\caption{Left panel: $(a_SX_S)^2$ for $S = t_0$, $N$, $w_0$,
         $\rho$ and $\pi$ along the unitary line,
         from the symmetric point $\delta m_l = 0$ down
         to the physical point
         $\delta m_l^* = (\delta m_u^* + \delta m_d^*)/2$
         (vertical dashed line) together with constant fits
         (for $\beta =5.50$, $\kappa_0 =0.12090$).
         Right panel: Values of $a_S^2$ for $S = \pi$, $N$, $\rho$,
         $w_0$ and $t_0$ using eq.~(\protect\ref{aS2}) for 
         $\kappa_0$ values on the symmetric line from
         $\kappa_0 = 0.12090$ to $\kappa_0 = 0.12099$,
         for $\beta = 5.50$, together with quadratic fits.
         The crossing of the horizontal and vertical dashed
         lines and circle gives an estimate for the common
         scale.}

\label{X2_scale}
\end{figure}

We shall now briefly mention here our progress in defining
and determining the scale using singlet quantities, collectively
denoted here by $X_S$. There are many possibilities such as pure
gluon quantities like the $r_0$ Sommer scale: $X_{r_0} = 1/r_0$,
or the $\sqrt{t_0}$ \cite{luscher10a}, $w_0$ \cite{borsanyi12a}
scales based on the Wilson gauge action flow: $X_{t_0} = 1/\sqrt{t_0}$,
$X_{w_0} = 1/w_0$, or quantities constructed using fermions.
One simple possibility in this case is to take the
`centre of mass' of the hadron octet. In all these cases
it can easily be shown that linear terms in $\delta m_q$ are absent,
\cite{bietenholz11a}. We then have from eqs.~(\ref{M2ps_expan}),
(\ref{M2N_expan}) in the unitary limit
(with $\overline{\delta m^2} = (\delta m_u^2 + \delta m_d^2 + \delta m_s^2)/3$)
\begin{eqnarray}
   X_{\pi}^2 = \sixth( M_{K^+}^2 + M_{K^0}^2 + M_{\pi^+}^2 
                                + M_{\pi^-}^2 + M_{\overline{K}^0}^2 + M_{K^-}^2)
            = M_{0\pi}^2 
                 + \left( \half\beta_0 + 2\beta_1 + 3\beta_2 \right)
                             \overline{\delta m^2} + \ldots \,,
\end{eqnarray}
and
\begin{eqnarray}
   X_N^2 = \sixth( M_p^2 + M_n^2 + M_{\Sigma^+}^2 +  M_{\Sigma^-}^2
                                  + M_{\Xi^0}^2 + M_{\Xi^-}^2 ) 
        = M_{0N}^2 + \half(B_0+B_1+B_3)\overline{\delta m^2} + \ldots \,.
\end{eqnarray}
In the left panel of Fig.~\ref{X2_scale} we plot various singlet
quantities $(a_SX_S)^2$ for $S = t_0$, $w_0$, $\pi$, $\rho$, $N$.
It is apparent that the constancy $(a_SX_S)^2$ holds over the
complete range from the symmetric point down to the physical point.
Using this enables us to use  $X_S^{\exp}$ to determine
the lattice spacing by
\begin{eqnarray}
   a_S^2 = {(a_SX_S)^2 \over X_S^{\exp\,2}} \,.
\label{aS2}
\end{eqnarray}
We shall define our lattice spacing here using $S=N$: $a_N$.
Of course depending on how well we have chosen our initial
$\kappa_0$ point, using a different singlet quantity (i.e.\ $S \not= N$)
will give a slightly different lattice spacing.
More ambitiously we can vary $\kappa_0$ to try to find a point
where we have a common scale. We have initiated a programme
to investigate this. In the right panel of Fig.~\ref{X2_scale}
we plot $a_S^2$ again for $S = \pi$, $N$, $\rho$, $w_0$, $t_0$
against various $\kappa_0$. The crossing of the $a_S^2$s
for $S = \pi$, $N$ and $\rho$ give an estimation of the
common scale as $a \approx 0.074(2)\,\mbox{fm}$.
We can now adjust $X_{t_0}^{\exp}$, $X_{w_0}^{\exp}$ to also
cross at this point to find a preliminary estimate
for these `intermediate scales' of
$\sqrt{t_0}^{\exp} \approx 0.153(7)\,\mbox{fm}$,
$w_0^{\exp} \approx 0.179(6)\,\mbox{fm}$.

Practically it is numerically advantageous to form dimensionless
ratios (within a multiplet): $\widetilde{M}^2 \equiv M^2 / X^2_S$
and re-write eqs.~(\ref{M2ps_expan}), (\ref{M2N_expan}) in
terms of $\widetilde{\alpha} \equiv \alpha/M_{0\pi}^2$, $\ldots$ and
$\widetilde{A}_i \equiv A_i / M_{0N}^2$, $\ldots$ in the
expansions. About $\sim O(80)$ PQ and unitary masses 
are used to determine these expansion coefficients and hence
the `physical' quark masses $\delta m_u^*$, $\delta m_d^*$, $\delta m_s^*$
and $\delta \mu_c^*$ as described in section~\ref{method}.


\section{Results and Conclusions}

\begin{figure}[h]

\begin{minipage}{0.45\textwidth}

      \begin{center}
         \includegraphics[width=7.00cm]
            {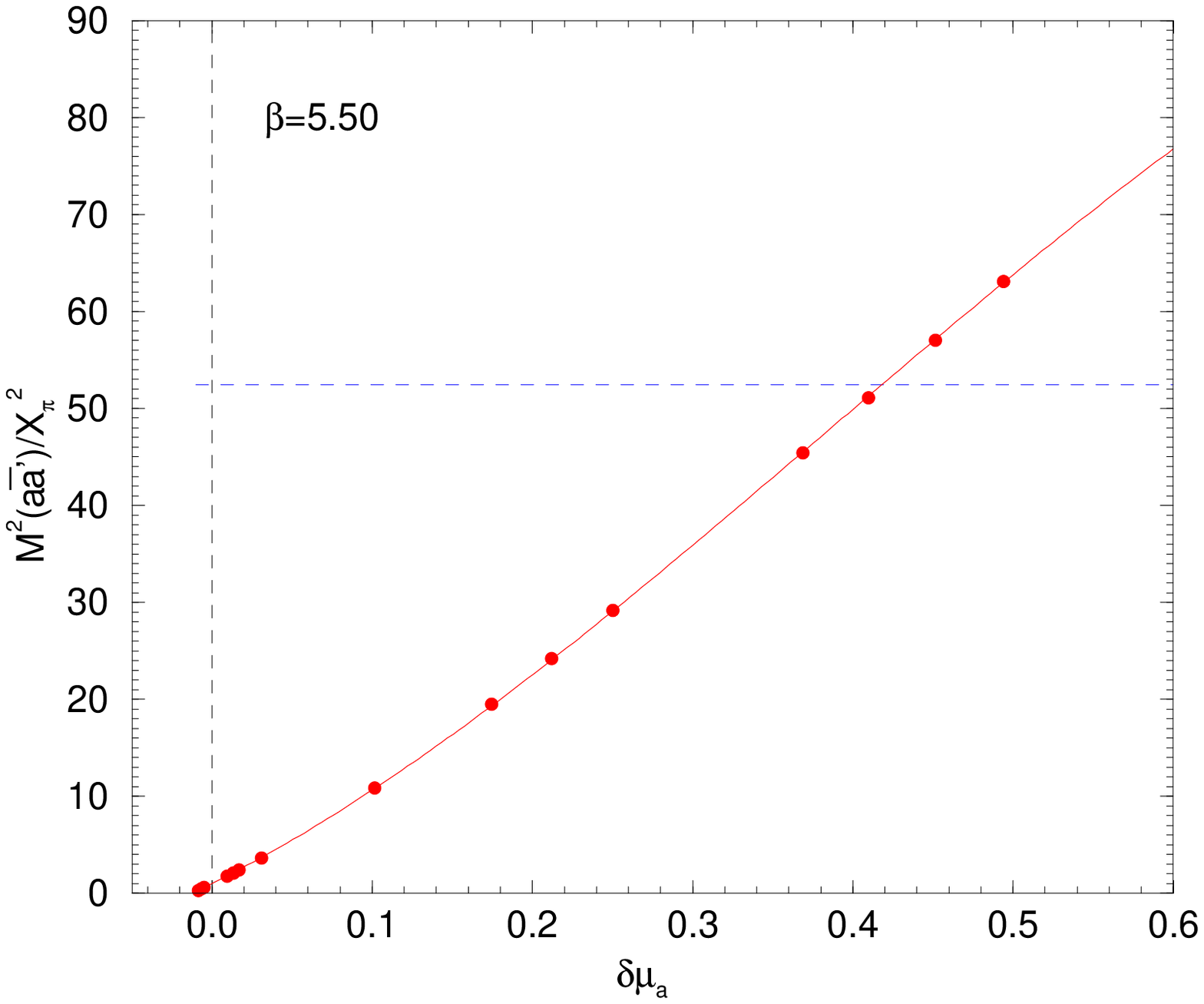}
      \end{center} 

\end{minipage}\hspace*{0.05\textwidth}
\begin{minipage}{0.45\textwidth}

      \begin{center}
         \includegraphics[width=7.00cm]
            {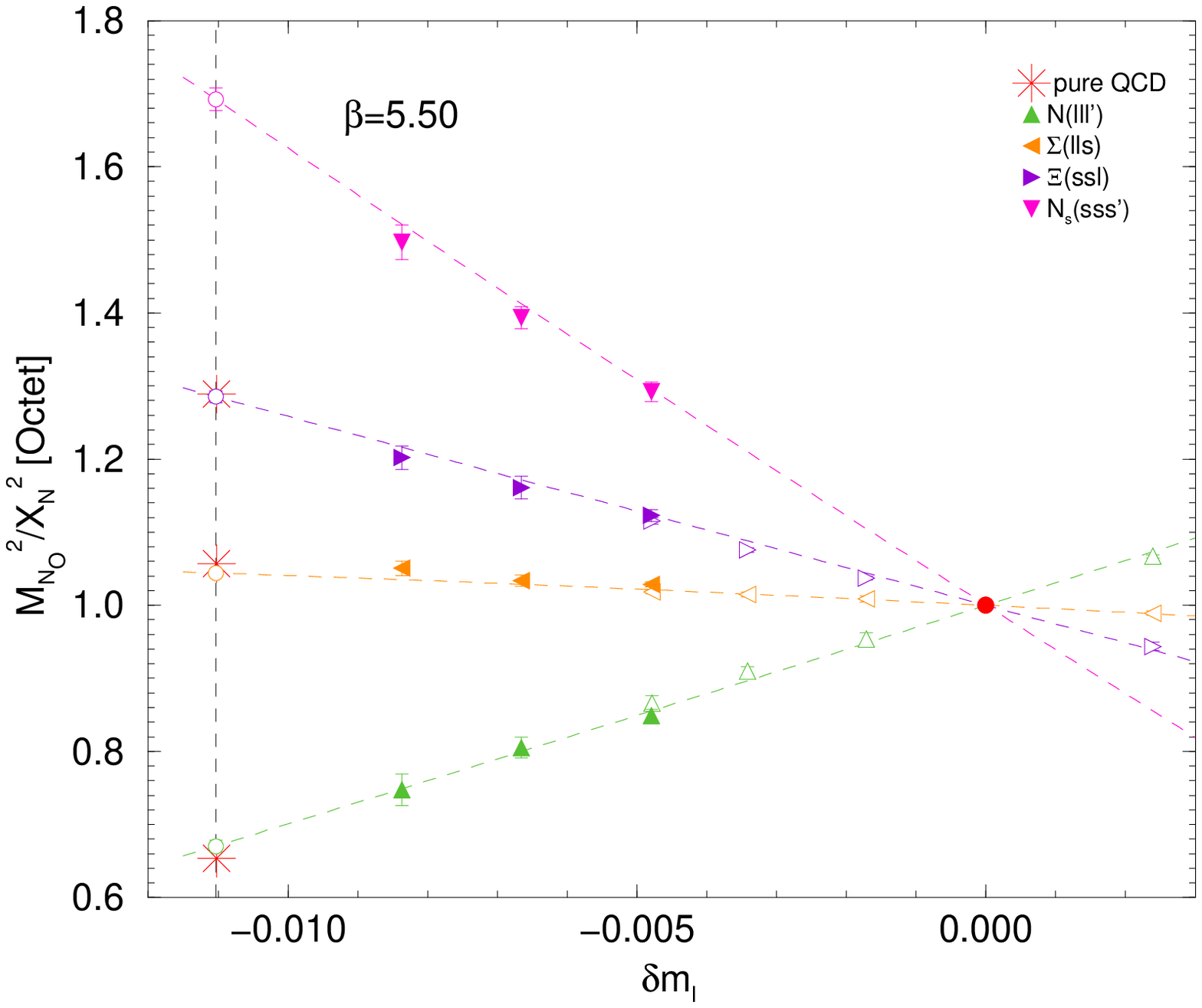}
      \end{center} 

\end{minipage}

\caption{Left panel:
         $\widetilde{M}^2(a\overline{a}^\prime)=M^2(a\overline{a}^\prime)/X_\pi^2$
         versus $\delta\mu_a$ for $\beta = 5.50$, together with
         the fit from eq.~(\protect\ref{M2ps_expan}). The vertical
         dashed line represents the symmetric point,
         while the horizontal dashed line
         is the physical value of $\widetilde{M}_{\eta_c}^2$. 
         Right panel: `fan' plot for the baryon octet,
         from the symmetric point $\delta m_l = 0$ to the physical point
         $\delta m_l^* = (\delta m_u^* + \delta m_d^*)/2$
         (vertical dashed line and stars) for $\beta = 5.50$.
         The filled triangles are from $32^3\times 64$ sized
         lattices, while the open triangles are from $24^3\times 48$
         sized lattices (not used in the fits). 
         The fits are again given from eq.~(\protect\ref{M2ps_expan}).} 

\label{fan_M2aa}
\end{figure}

We now discuss our results. In Fig.~\ref{fan_M2aa}
left panel, we show the diagonal pseudoscalar mesons
$\widetilde{M}^2(a\overline{a}^\prime)$ (to avoid a three dimensional
plot) versus $\delta\mu_a$ together with the fit from
eq.~(\ref{M2ps_expan}) (using the prime notation $a^\prime$ to mean
a distinct quark from $a$ but degenerate in mass).
The horizontal dashed line represents the physical value
of $\widetilde{M}_{\eta_c}^2$, the intersection with the fit curve
gives a determination of $\delta \mu_c^*$. In the
right panel we show a `fan' plot of $\widetilde{M}_N^2$,
$\widetilde{M}_{\Sigma}^2$, $\widetilde{M}_{\Xi}^2$ and $\widetilde{M}_{N_s}^2$
against $\delta m_l$, together with the fit using eq.~(\ref{M2ps_expan}).
Note that the scales involved are rather different, for the
unitary masses $|\delta m_l| \sim 0.01$ and the LO terms in
eq.~(\ref{M2ps_expan}) or (\ref{M2N_expan}) dominate,
\cite{bietenholz11a}, while for the PQ masses (reaching up to
the charm masses) we have $\delta\mu_a \sim 0.4$ but still 
with rather moderate curvature.

\begin{figure}[h]

\begin{minipage}{0.45\textwidth}

      \begin{center}
         \includegraphics[width=7.00cm]
            {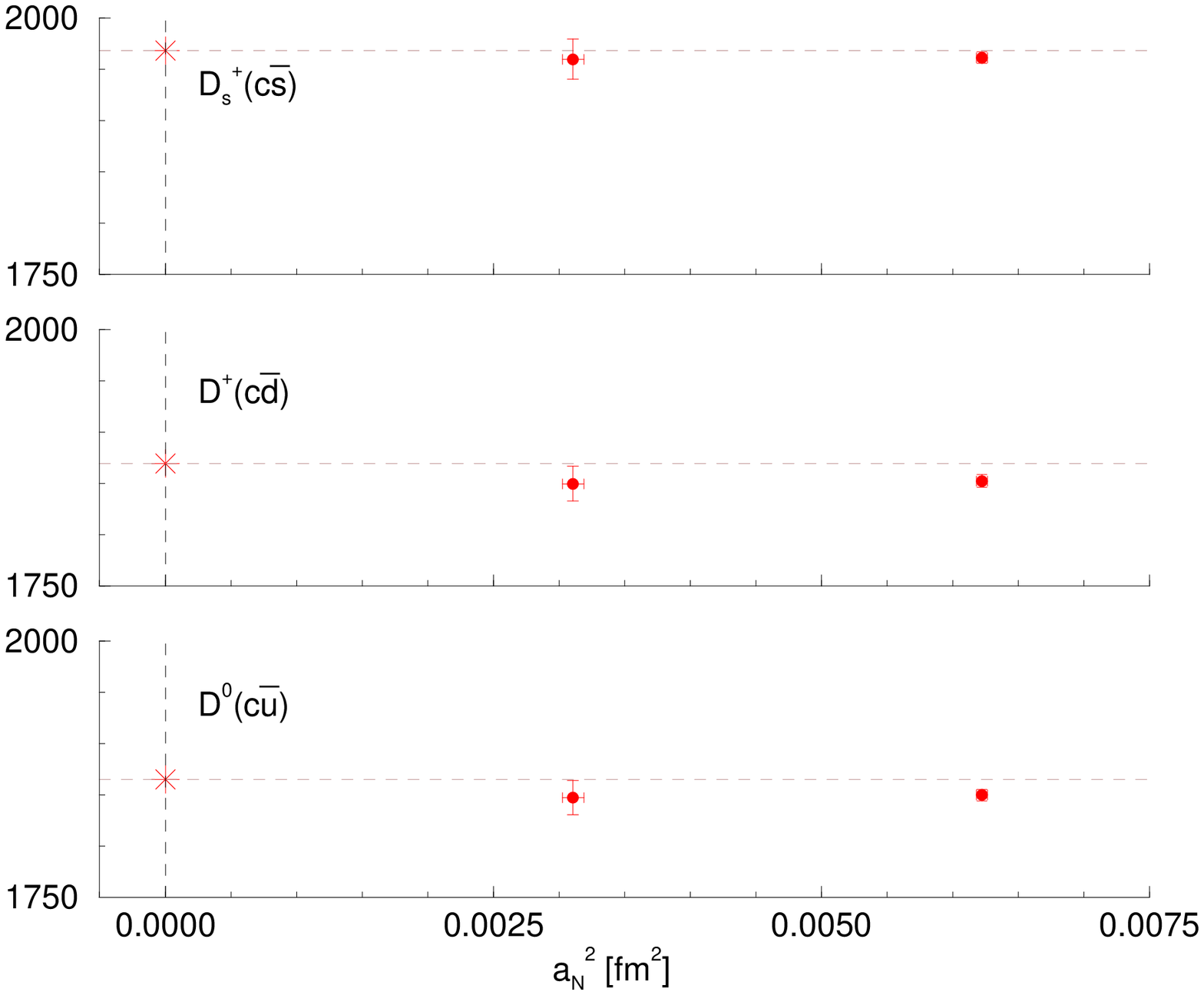}
      \end{center} 

\end{minipage}\hspace*{0.05\textwidth}
\begin{minipage}{0.45\textwidth}

      \begin{center}
         \includegraphics[width=7.00cm]
            {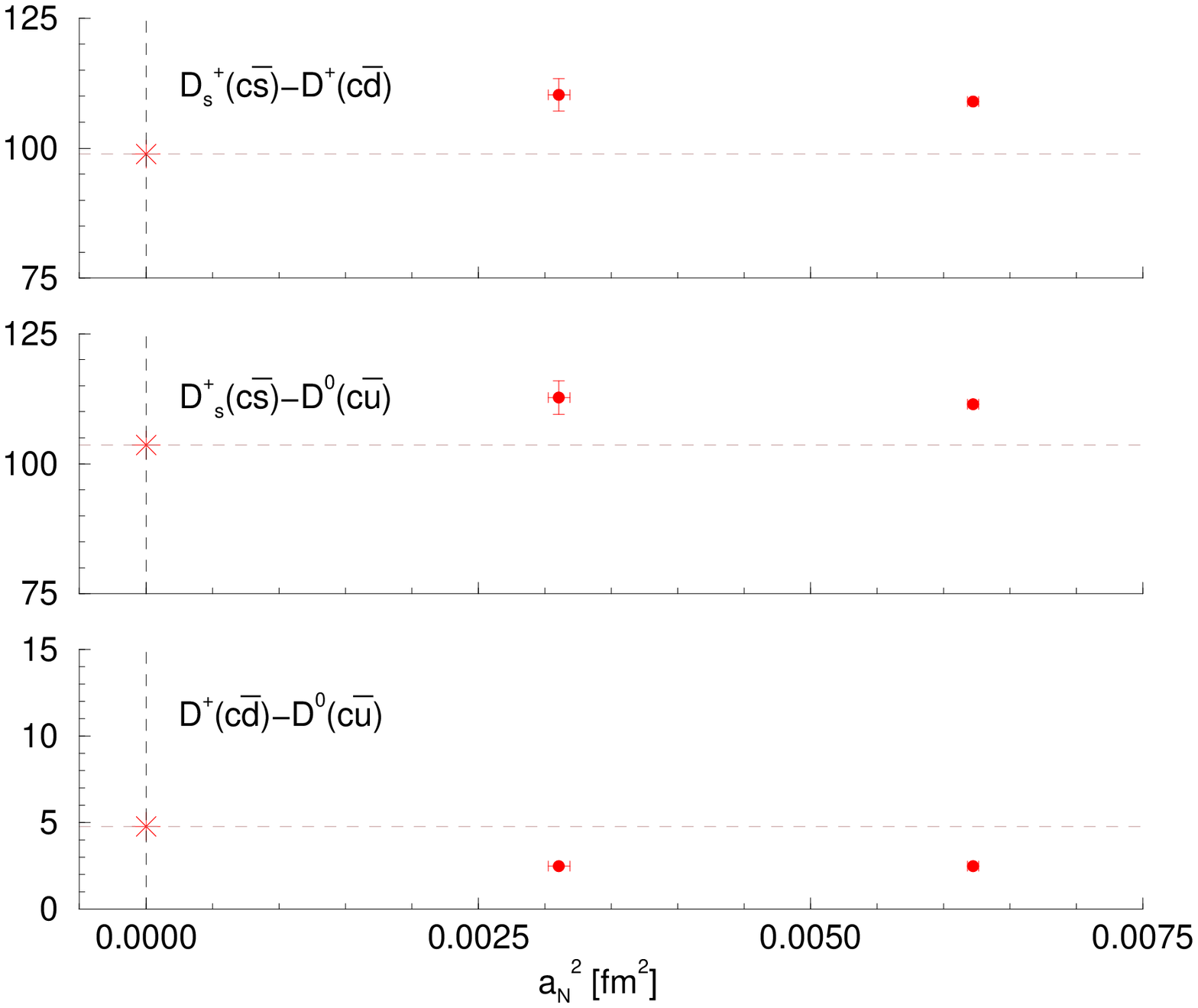}
      \end{center} 

\end{minipage}

\caption{Left panel: $D^0(c\overline{u})$, $D^+(c\overline{d})$
         and $D_s^+(c\overline{s})$.
         Right panel: $D^+(c\overline{d})-D^0(c\overline{u})$,
         $D_s^+(c\overline{s})-D^0(c\overline{u})$ and
         $D_s^+(c\overline{s})-D^+(c\overline{d})$
         mass splittings. (All values in $\mbox{MeV}$.)
         The experimental values are given as red stars.
         To guide the eye, we extend these values as
         horizontal dashed lines.}

\label{D0_DD_pD_ps}
\end{figure}

In Fig.~\ref{D0_DD_pD_ps} we show $D^0(c\overline{d})$,  
$D^+(c\overline{d})$ and $D_s^+(c\overline{s})$ against our
$a_N^2$ lattice spacings (left panel) and their mass differences
(right panel). These mass differences in particular are sensitive
to unknown QED effects (the present computation is for
pure QCD only).  As we currently have only two  lattice
spacings (and are also increasing their statistics)
the results are to be regarded as preliminary
and we do not presently attempt a continuum extrapolation.
However there do not seem to be strong scaling violations
present.

\begin{figure}[h]

\begin{minipage}{0.45\textwidth}

      \begin{center}
         \includegraphics[width=7.00cm]
            {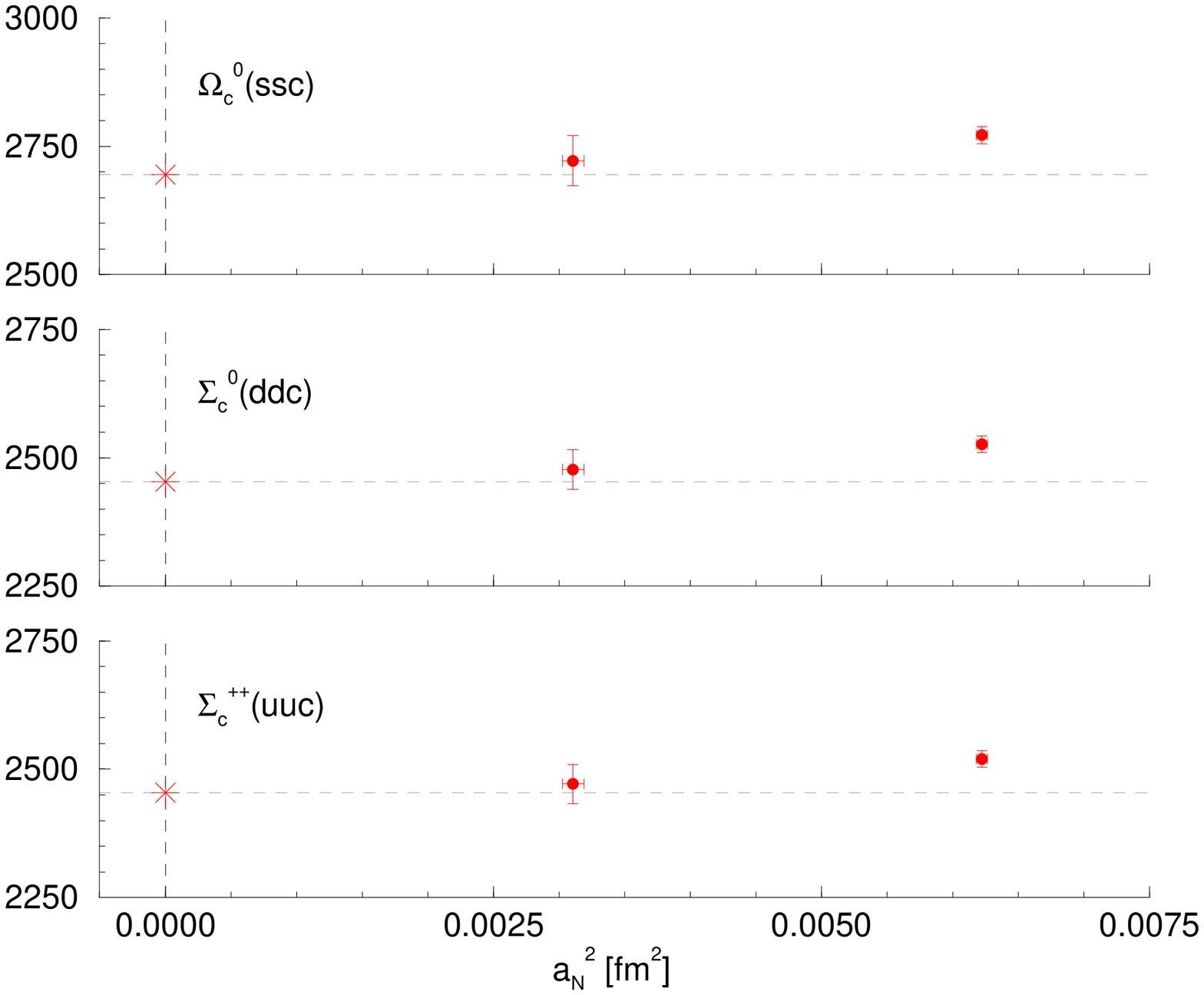}
      \end{center} 

\end{minipage}\hspace*{0.05\textwidth}
\begin{minipage}{0.45\textwidth}

      \begin{center}
         \includegraphics[width=7.00cm]
            {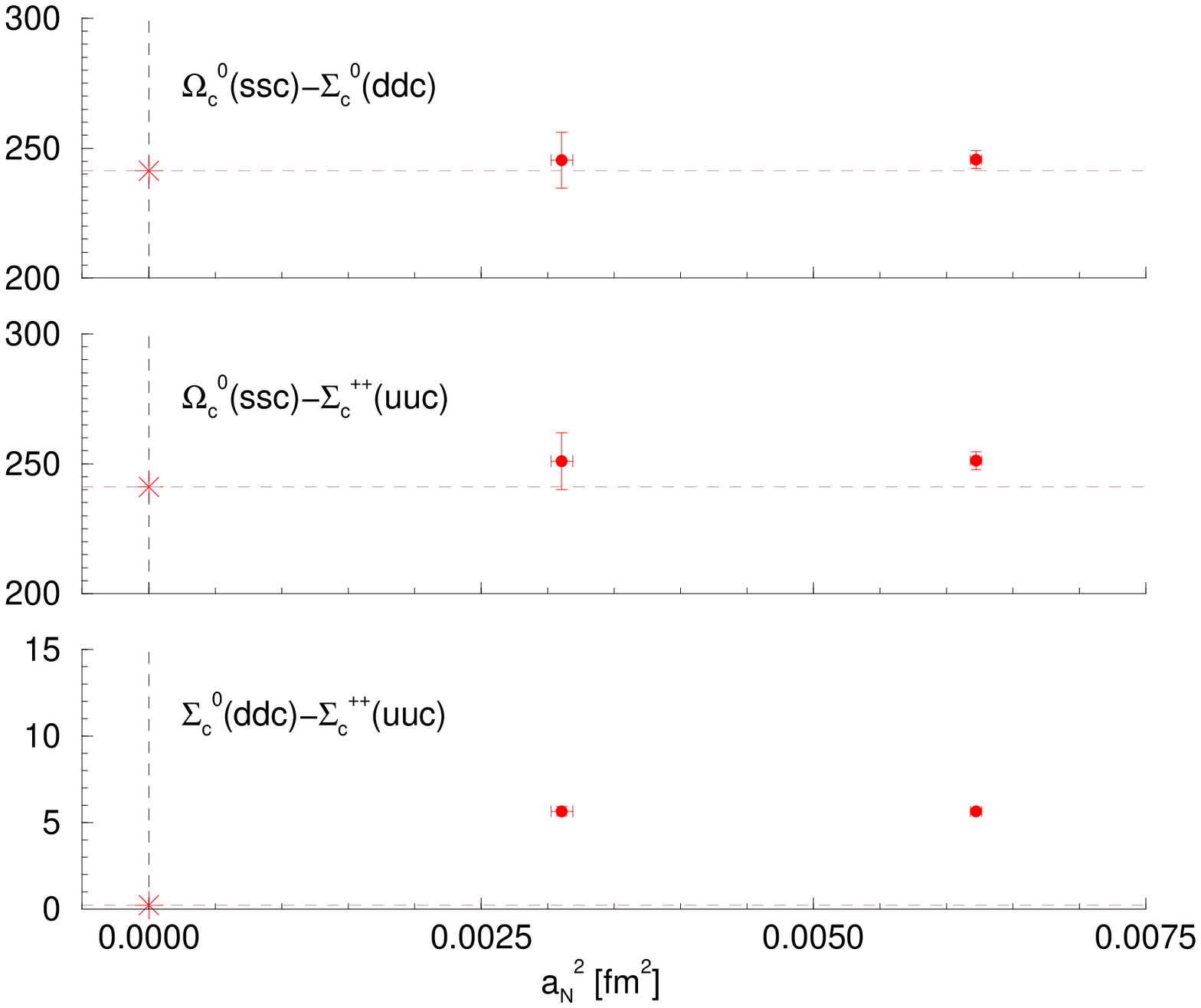}
      \end{center} 

\end{minipage}

\caption{Left panel: $\Sigma_c^{++}(uuc)$, $\Sigma_c^0(ddc)$,
        $\Omega_c^0(ssc)$.
        Right panel: $\Sigma_c^0(ddc) - \Sigma_c^{++}(uuc)$,
        $\Omega_c^0(ssc) - \Sigma_c^{++}(uuc)$,
        $\Omega_c^0(ssc) - \Sigma_c^0(ddc)$ mass splittings.}

\label{Ceq1_baryon}
\end{figure}

\begin{figure}[h]

\begin{minipage}{0.45\textwidth}

      \begin{center}
         \includegraphics[width=7.00cm]
            {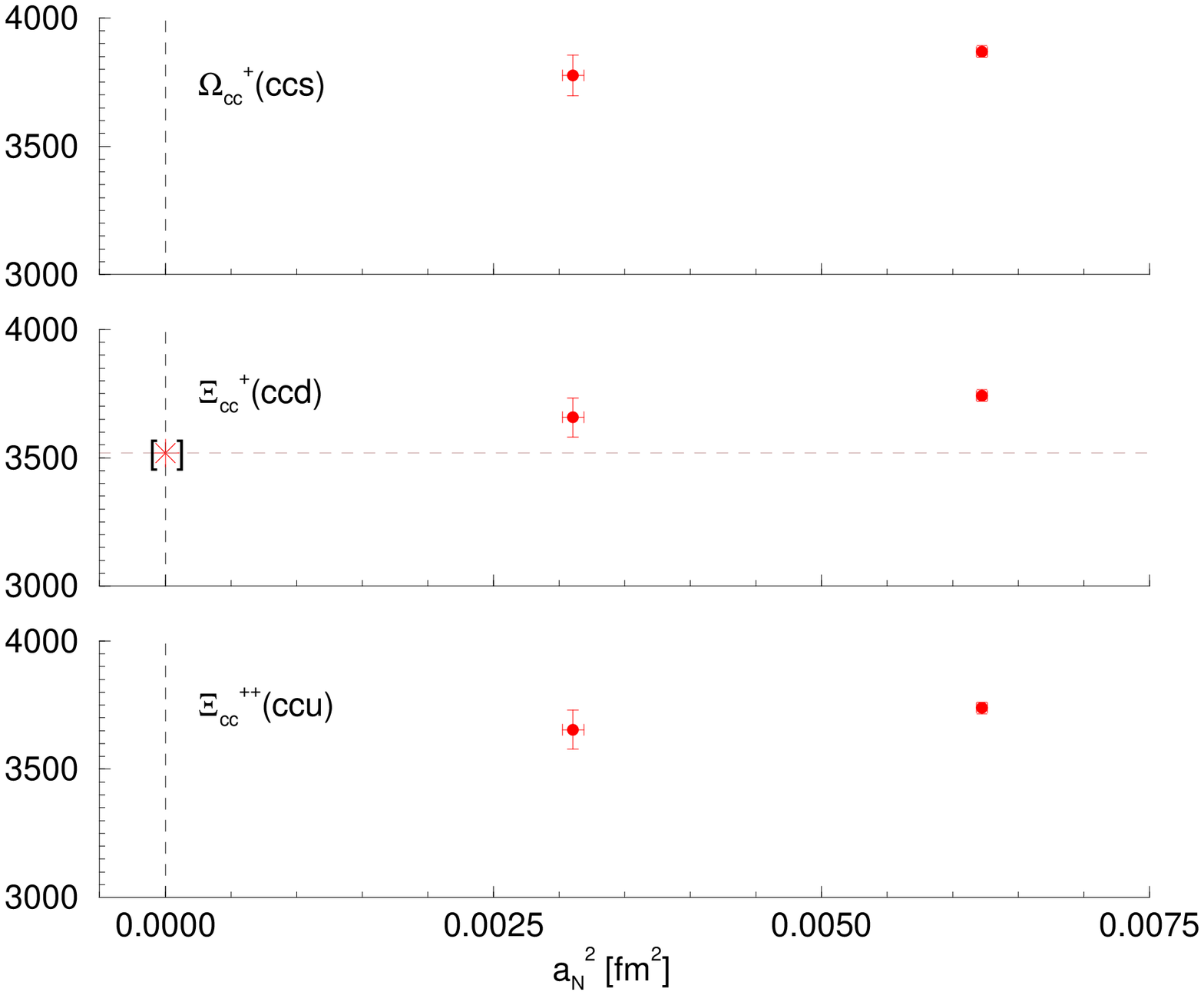}
      \end{center} 

\end{minipage}\hspace*{0.05\textwidth}
\begin{minipage}{0.45\textwidth}

      \begin{center}
         \includegraphics[width=7.00cm]
            {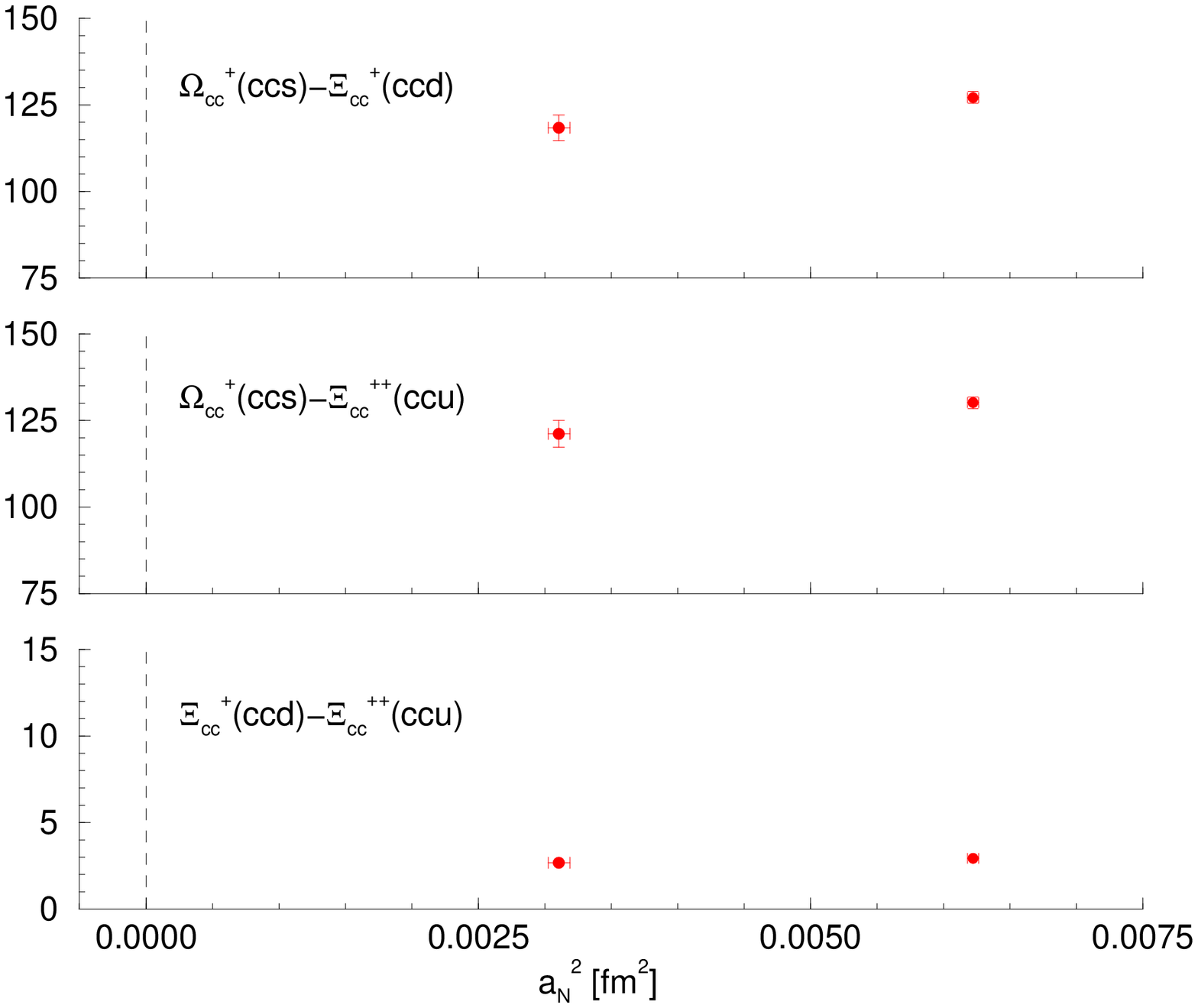}
      \end{center} 

\end{minipage}

\caption{Left panel: $\Xi_{cc}^+(ccd) - \Xi_{cc}^{++}(ccu)$.
        The SELEX Collaboration result,
        \protect\cite{selex02a}, for $\Xi_{cc}^+$ is shown as a red star.
         Right panel: $\Omega_{cc}^+(ccs) - \Xi_{cc}^{++}(ccu)$,
         $\Omega_{cc}^+(ccs) - \Xi_{cc}^+(ccd)$ mass splittings.}

\label{Ceq2_baryon}
\end{figure}

In Figs.~\ref{Ceq1_baryon} and \ref{Ceq2_baryon}
we show the $C = 1$ and $C=2$ charmed baryons (from the spin $1/2$
$20$-plet). Again while we do not see significant lattice effects
in either case, we gain an impression that those present are
a little larger for the doubly charmed baryons than for the
singly charmed mesons.

In conclusion we note that we have developed a method to 
determine some open charm states using a precise $SU(3)$
flavour symmetry breaking expansion enabling  $u$, $d$, $s$
quarks to approach the physical point while the $c$ quark is
treated as PQ. The expansion appears to be highly convergent.
The method can be extended to other states. In a $2+1$ world there
is no $\Sigma^0$ - $\Lambda^0$ mixing, but the determined
coefficients can be used to compute $\Sigma^0(uds)$ - $\Lambda^0(uds)$
mixing, \cite{qcdsf14a}. Therefore computing e.g.\
$\Sigma_c^+$ - $\Lambda_c^+$, $\Xi_c^0$ - $\Xi_c^{\prime 0}$
mixing is possible. Furthermore the method can be applied
to the baryon decuplet and QED effects can be introduced,
\cite{schierholz13a}.


\section*{Acknowledgements}


The numerical configuration generation (using the BQCD lattice
QCD program \cite{nakamura10a}) and data analysis 
(using the Chroma software library) was carried out
on the IBM BlueGene/Q using DIRAC 2 resources (EPCC, Edinburgh, UK),
the BlueGene/P and Q at NIC (J\"ulich, Germany), the
SGI ICE 8200 at HLRN (Berlin--Hannover, Germany)
and on the NCI National Facility in Canberra, Australia
(supported by the Australian Commonwealth Government).
The BlueGene/P codes were optimised using Bagel.
This investigation has been supported partly by the DFG
under contract SFB/TR 55 (Hadron Physics from Lattice QCD)
and by the EU grants 227431 (Hadron Physics2), 283826 (Hadron Physics3)
and 238353 (ITN STRONGnet). JN was partially supported by EU grant
228398 (HPC-EUROPA2). JMZ is supported by the Australian Research
Council grant FT100100005. We thank all funding agencies.



\end{document}